\documentclass[]{aastex631}
\makeatletter
\let\frontmatter@title@above=\relax
\makeatother
\usepackage{lipsum}

\graphicspath{{./}{figures/}}
\usepackage{multirow}
\usepackage{xcolor}
\usepackage[normalem]{ulem}
\usepackage{natbib}
\begin{document}

\title{New Insights into Variations in Enceladus Plume Particle Launch Velocities from Cassini-VIMS spectral data}

\author[0000-0002-0803-2678]{H. Sharma}
\affiliation{University of Idaho \\
875 Perimeter Dr \\
Moscow, ID 83844}

\author[0000-0002-8592-0812]{M. M. Hedman}
\affiliation{University of Idaho \\
875 Perimeter Dr \\
Moscow, ID 83844}

\author{S. Vahidinia}
\affiliation{NASA Ames Research Center \\
Mountain View, CA 94035}



\begin{abstract}
Enceladus' plume consists mainly of a mixture of water vapor and solid ice particles that may originate from a subsurface ocean. The physical processes underlying Enceladus’ plume particle dynamics are still being debated, and quantifying the particles' size distribution and launch velocities can help constrain these processes. Cassini's Visual and Infrared Mapping Spectrometer (VIMS) observed the Enceladus plume over a wavelength range of 0.9 $\mu$m to 5.0 $\mu$m for a significant fraction of Enceladus' orbital period on three dates in the summer of 2017. We find that the relative brightness of the plume on these different dates varies with wavelength, implying that the particle size distribution in the plume changes over time. These observations also enable us to study how the particles' launch velocities vary with time and observed wavelength. We find that the typical launch velocity of particles remains between 140 m$s^{-1}$ and 148 m$s^{-1}$ at wavelengths between 1.2 $\mu$m and 3.7 $\mu$m. This may not be consistent with prior models where particles are only accelerated by interactions with the vent walls and gas, and could imply that mutual particle collisions close to the vent are more important than previously recognized.
\end{abstract}


\section{Introduction}
\label{sec:intro}

Enceladus emits a plume of water vapor and icy particles from a series of fissures located near its south pole \citep{spencer2006cassini,dougherty2006identification,porco2006cassini}. This plume can shed light on the processes operating inside Enceladus and the driving forces behind this geological activity. Measurements from several instruments onboard the Cassini spacecraft have been utilized to examine the plume's characteristics \citep{hansen2006enceladus,waite2006cassini,spahn2006cassini,waite2009liquid,hedman2009spectral,postberg2009sodium,schenk2018enceladus}. At the same time, several different theoretical models have been developed to explain various aspects of the plume \citep{kieffer2006clathrate,hurford2007eruptions,schmidt2008slow,brilliantov2008geysers,ingersoll2010subsurface,goldstein2018enceladus, 2016PNAS..113.3972K, 2021PSJ.....2..157E}.

One important source of information about the physics behind Enceladus' plume are its variations over time periods ranging from days to years. Variations have been observed in plume's total particle output by both the Visual and Infrared Mapping Spectrometer (VIMS) instrument \citep{hedman2013observed} and the Imaging Science Subsystem (ISS) cameras \citep{nimmo2014tidally,porco2014geysers,2015AJ....150...96H,ingersoll2017decadal,ingersoll2020time} onboard the Cassini spacecraft. Most dramatically, both the ISS and VIMS data show that the plume's ice grain output varies systematically with the moon's orbital phase (the angular distance between the moon's current position and its orbital pericenter, equivalent to the moon's orbital mean anomaly). The primary maximum in the plume's brightness occurs close to orbital apocenter (orbital phase of 180$^\circ$), where the particle output is roughly four times higher than it is at other points in its orbit. The most likely cause of the variations on orbital timescales are the tidal stresses experienced by Enceladus as it moves in its eccentric orbit around Saturn \citep{2007Natur.447..292H,2012Icar..220..896H,nimmo2007shear,smith2008tidally,goldstein2018enceladus}. Variations in the plume's activity on longer time scales were also seen by both VIMS \citep{hedman2013observed} and ISS
\citep{ingersoll2017decadal,2018LPI....49.2003P,ingersoll2020time} which may be due to either a 5$\%$ decrease in the eccentricity of the orbit as part of a $\sim$ 11-year tidal cycle, or slow (and perhaps seasonal) changes in the clogging of vents  \citep{2018LPI....49.2003P,ingersoll2017decadal}. 
The plume is also prone to stochastic time variability on month-to-year timescales \citep{ingersoll2020time} whose origins are still unclear and maybe due to individual jets turning on and off \citep{2007Natur.449..695S, ingersoll2010subsurface,2012Icar..220..896H, nimmo2014tidally,  porco2014geysers, nakajima2016controlled, teolis2017enceladus}.

Compared to the relatively dramatic brightness variations listed above,
trends in other plume particle properties like launch velocity and size distributions are more subtle.
\citet{hedman2009spectral} found some differences in the spectral properties of the plume among the early VIMS observations, but these were of marginal statistical significance.
Both \citet{hedman2013observed} and \citet{ingersoll2017decadal} found small variations in the launch velocity with orbital phase.
\citet{nimmo2014tidally}
also reported an essentially constant scale height parameter for the plume. These observations are generally consistent with models where increasing crack width increases the total mass flow but has little effect on particle velocities  \citep{ingersoll2010subsurface}.

\begin{table}[t]
\caption{The three data sets studied in this paper.} 
\footnotesize
\centering
\begin{tabular}{cccccccc} 
\hline\hline
 & & & & & & No. of & No. of\\ 
Orbit no. & Orbital & Phase & Range\footnote{Range defines the distance between Enceladus and Cassini.} & Longitude\footnote{Sub-spacecraft longitude on Enceladus} & Cubes &  cubes & cubes \\
Date & Phase\footnote{The orbital phase refers to the position of Enceladus in its orbit around Saturn, also known as mean anomaly} & angle\footnote{The phase angle is the angle formed between the Sun, the target being imaged (Enceladus and its plume), and the spacecraft (Cassini). Note that this number does not increase continuously over the course of the observation, but stays in the given range.} & (km) & & & before & after \\
& & & & & & binning & binning \\
\hline
279/June 18th, 2017 & 162 - 319$^\circ$ & 156 - 161$^\circ$ & 8,37,300 - 10,93,300 & 11 - 164$^\circ$ & CM$\_$1876443559$\_$1 - CM$\_$1876495215$\_$1 & 302 & 28\\
286/Aug 2nd, 2017 & 162 - 286$^\circ$ & 159 - 161$^\circ$ & 8,22,700 - 10,63,000 & 29 - 150$^\circ$ & CM$\_$1880355368$\_$1 - CM$\_$1880396071$\_$6 & 293 & 26\\
290/Aug 28th, 2017 & 130 - 281$^\circ$ & 156 - 158$^\circ$ & 8,68,600 - 11,40,800 & 13 - 160$^\circ$ & CM$\_$1882597042$\_$1 - 
CM$\_$1882646878$\_$5 & 240 & 23\\
\hline 
\end{tabular}
\label{tab:obs} 
\end{table}

\begin{figure}[ht]
\plotone{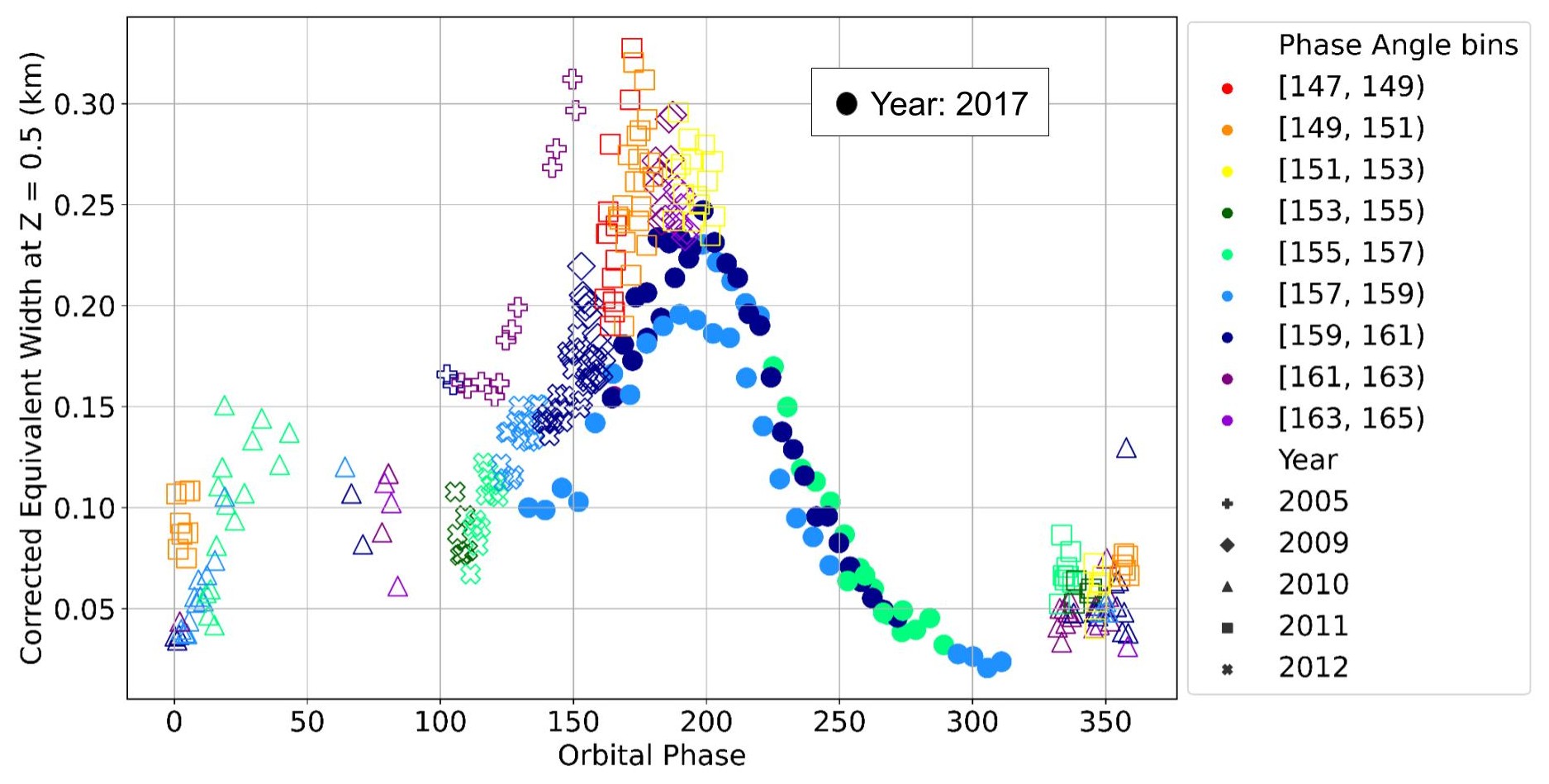}
\caption{Overview of variations in the plume's brightness (expressed in terms of the corrected Equivalent Width at a wavelength of 0.88 - 1.56 $\mu$m and an altitude of 85 km, see text for details) as a function of orbital phase. This figure compares the plume brightness estimates derived in this study (shown as solid dots) with previously published estimates from earlier in the Cassini mission \citep[shown as empty symbols][]{hedman2013observed}. Note that the corrected Equivalent Width used in this particular plot includes the phase angle correction described in \cite{hedman2013observed} in order to facilitate comparisons among the different data sets.
\label{fig:combined}}
\end{figure}

This paper aims to quantify variations in the plume's properties
using spectral data obtained by the Visual and Infrared Mapping Spectrometer (VIMS) onboard the Cassini spacecraft \citep{brown2004cassini}. VIMS can provide new information about trends in the particle size and velocity distribution over time because it was able to observe the plume over a broad range of near-infrared wavelengths. More specifically, we will examine VIMS data obtained on three dates -- June 18th, Aug 2nd and Aug 28th, 2017.  There is a clear maxima in the plume's brightness around apoapsis in the VIMS data on all three of these dates (see Figure~\ref{fig:combined}), consistent with previous ISS results \citep{ingersoll2020time}.
The VIMS observations at short wavelengths also show the same variations in the plume's brightness among the three orbits, where the brightness increases from June 18th to Aug 2nd and decreases again on Aug 28th as reported by \citet{ingersoll2020time}. However, the VIMS data show that these variations in plume brightness across the three different orbits being studied here are not the same at all wavelengths, suggesting the plume's particle size distribution also varied over this time period of 3 months. In addition, we find that the launch velocities of the plume particles do not vary as much with wavelength as published models would predict \citep{schmidt2008slow, degruyter2011cryoclastic} but see \citet{2014AGUFM.P51F..05S} for modified models that may be consistent with these observations.

Our methods for extracting information about the plume from the VIMS data are described in Section~\ref{sec:method}. This analysis begins by converting the raw VIMS data to estimates of the plume's spectra at various distances from the south pole. These spectra are then converted into constraints on the plume's overall intensity and the particles' typical launch velocity assuming the particles follow ballistic trajectories. The results of these calculations are presented in Section~\ref{sec:results}. Finally, in Section~\ref{sec:discuss} we further discuss the implication of the observed spectral trends in the brightness and typical launch velocity of particles. Note that this paper is focused on documenting the spectral trends in these data. Detailed spectral modeling of these trends will be the subject of future work.

\section{Methods} \label{sec:method}
This section describes the steps by which the relevant VIMS data are reduced into standardized estimates of the plume brightness and typical launch velocity at a range of wavelengths. Section 2.1 describes the geometry and calibration of the VIMS observations for the 3 dates in 2017 being studied here. Section 2.2 details how these data are processed to obtain high signal-to-noise spectra of the plume at different altitudes. Finally, Section 2.3 shows how these data are fit to obtain the brightness and typical launch velocity at different wavelengths.

\subsection{Data} \label{subsec:obs}
The Visual and Infrared Mapping Spectrometer (VIMS)  was an imaging spectrometer onboard the Cassini spacecraft that covered the 0.3 $\mu$m - 5.1 $\mu$m wavelength range using 352 spectral channels. This instrument could view an array of up to 64 x 64 locations in the sky to produce a spectral-spatial image ``cube" \citep{brown2004cassini}. In this paper, we focus exclusively on the infrared spectra obtained by the VIMS-IR channel that measured the brightness in 256 wavelength bands between 0.88 $\mu$m and 5.1 $\mu$m with a typical spectral resolution of 0.016 $\mu$m. Further, we have removed the data corresponding to spectral channels at 1.23 $\mu$m (channel 118) and at 4.75 $\mu$m (channel 330) as they contain null values due to being hot pixels on the detector \citep{clark2018vims} and exclude data beyond 4.0 $\mu$m because the signal-to-noise ratio is significantly lower at these wavelengths. This reduces the number of spectral channels considered here from 256 to 186.

This investigation examines VIMS observations of Enceladus from 3 different Cassini orbits (designated 279, 286 and 290) corresponding to 3 days in 2017 - June 18th, Aug 2nd and Aug 28th. During all three of these days, VIMS viewed Enceladus from similarly high phase angles (156$^\circ$ - 162$^\circ$) over a similar range of the moon's orbital phase. The parameters for these three observations are listed in Table~\ref{tab:obs}. Note that all three observations cover orbital phases around 180$^\circ$, when the plume is most active, and are at high enough phase angles for the plume signals to be clearly detectable.

\begin{figure}[t]
\plotone{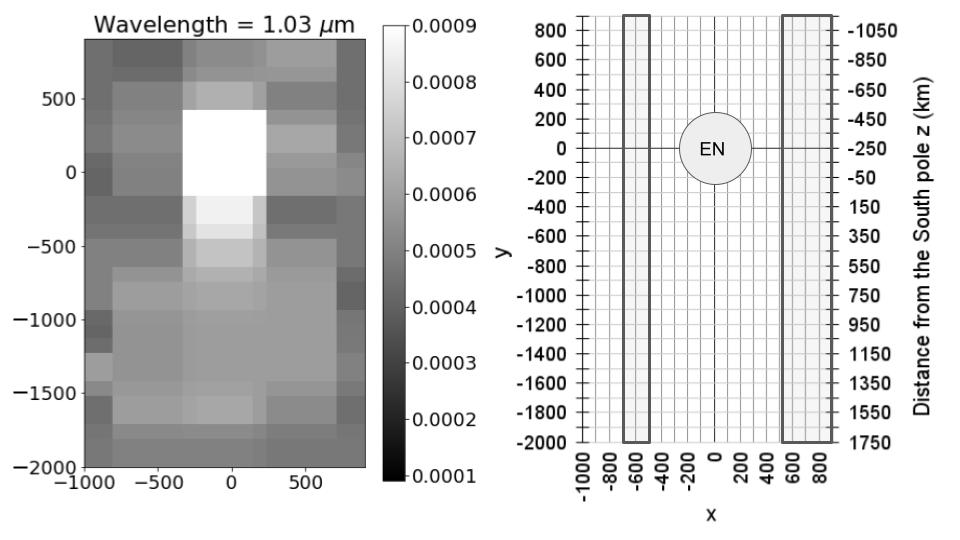}
\caption{The spatial geometry of the re-projected VIMS cubes. On the left is a sample image of Enceladus and its plume at a wavelength of 1.03 $\mu$m derived from a single re-projected cube (CM$\_$1876456410$\_$1) obtained at an orbital phase of 200$^\circ$ during Orbit no. 279 on June 18th. On the right is the geometry of the re-projected VIMS cubes. The  re-projected data plane is defined using cartesian co-ordinates [x, y], with Enceladus at the center at [0, 0] and the negative y axis is aligned with the moon's spin axis. The left vertical axis and the horizontal axis show the [x, y] coordinates for the cube. For the right vertical axis of the figure the y co-ordinate has been converted into altitude above the Enceladus' south pole ($z = -(250 + y)$ in km). The outlined regions on either side correspond to the region used to calculate background signal levels.
\label{fig:filter}}
\end{figure}

The raw data in each cube are converted into I/F values (a standard measure of reflectance) using standard calibration routines (the specific calibration being RC19 \citep{clark2018vims}). To facilitate comparisons among the observations, the observation geometry for each cube is computed using the appropriate SPICE kernels, and the brightness data are re-projected onto a regular array of cartesian co-ordinates [x, y] containing the moon's spin axis. In these coordinates the center of Enceladus is located at [0, 0] and the negative y-axis is aligned with Enceladus' spin axis. In the resulting maps x ranges from -1000 km to 900 km and y ranges from -2000 km to 900 km. Both coordinates are sampled every 100 km. Figure~\ref{fig:filter} shows the extent of the co-ordinates x and y. We also use the y coordinate to compute the distance from Enceladus' south pole (i.e. the plume's altitude) as  $z = -(250 + y)$ in km.

\subsection{Extracting Brightness Spectra at different altitudes and times} \label{subsec:filter}

The first step in extracting plume spectra from these cubes is to remove cubes that had instrumental artifacts that made their spectra discrepant from the rest of the observation. We identified these anomalous cubes by first averaging the brightness in the region of Enceladus' plume at each wavelength over all values of x from -400 km to 400 km and all values of y from -300 km and beyond. A median filter is then applied to this list of average brightness values at each wavelength. Any image/cube that lies outside the 3$\sigma$ range of the median brightness was flagged as an outlier for that spectral channel. We then compared the outlier list of each spectral channel and if a cube appeared as an outlier for over 35 spectral channels it was regarded as unreliable and so removed from further consideration. This procedure led to the removal of 22 cubes from the data on June 18th/Orbit no. 279, 24 cubes from the data on Aug 2nd/Orbit no. 286 and 11 cubes from the data on Aug 28th/Orbit no. 290 (see Appendix~\ref{cubes} for an explicit list of these cubes). 

After removing the outliers highlighted across orbital phase and wavelength and before applying background removal techniques the cubes are co-added.
Spectra derived from the individual remaining cubes had low signal-to-noise, so we averaged together sets of 10 cubes to improve the signal-to-noise in the spectra. Note that each of these sets of cubes corresponds to a relatively narrow range of orbital phases, so this averaging does not significantly affect our ability to quantify variations in plume activity. Table~\ref{tab:obs} shows the range of cubes corresponding to each orbit, the number of cubes before binning and after binning sets of 10 cubes each.

\begin{figure}[t]
\plotone{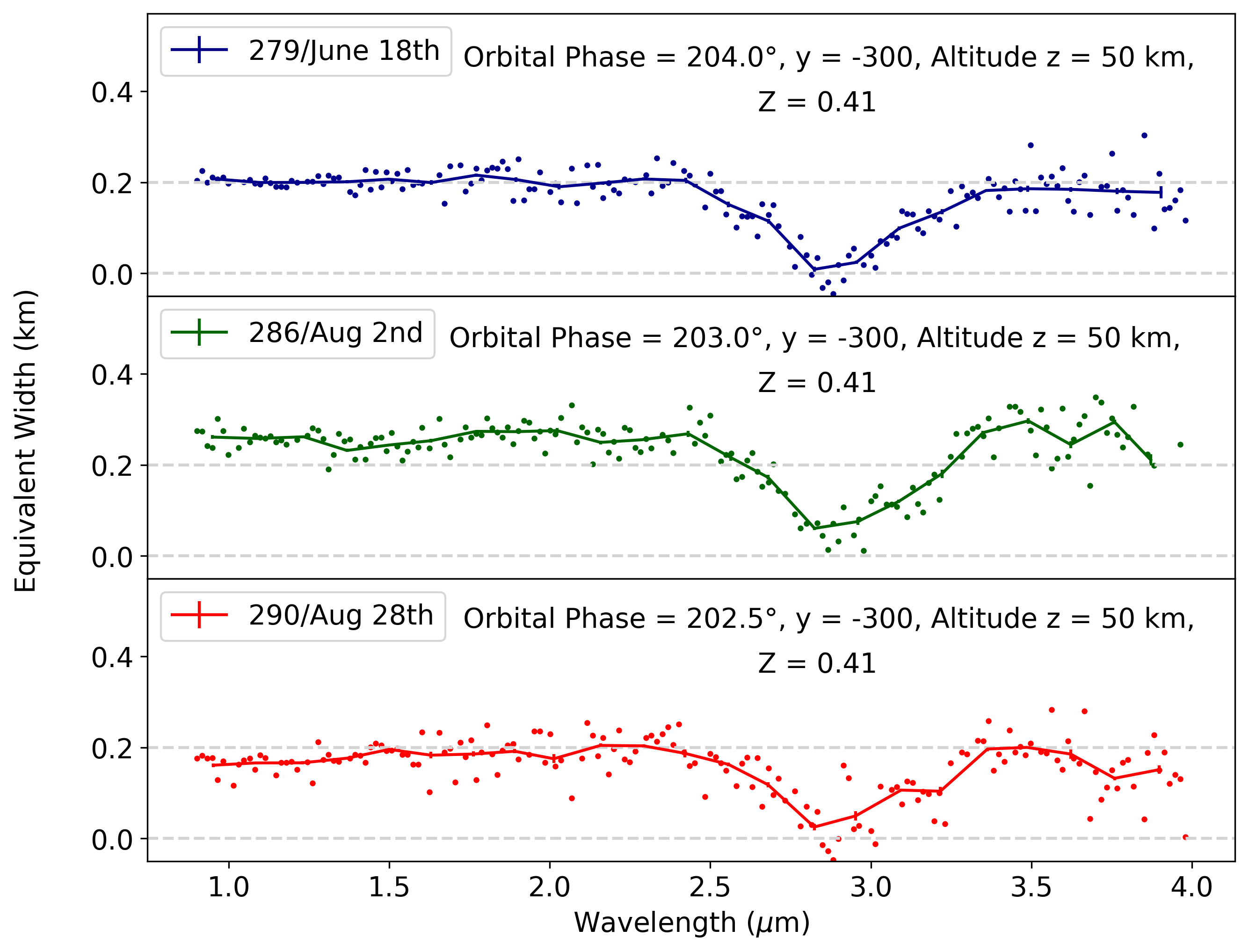}
\caption{Example Enceladus plume spectra, showing the plume's Equivalent Width (in km, see sub-section~\ref{subsec:filter}) versus wavelength (in $\mu$m) at similar orbital phases for all three orbits. An orbital phase close to 200$^\circ$ is chosen as the signal is strongest closer to the plume maxima at apoapsis. The points show the Equivalent Widths for individual spectral channels (after removing outliers before 2.5 $\mu$m and after 3.3 $\mu$m) and the line plot with error bars shows the data after averaging each spectrum over 8 wavelength channels. 
\label{fig:wavelength filter}}
\end{figure}

From each of these co-added cubes, we derive plume spectra as a function of altitude. Preliminary investigations of these data indicated that the majority of the plume signal exists between x = -400 km and x = 400 km. Any signal in the range x $<$ -400 km and x $>$ 400 km is therefore primarily due to background signals from the instrument or the E ring. We estimated this background by fitting the brightness values for -700 km $<$ x $<$ -400 km and x $>$ 400 km at each value of y to a linear trend. Note that the brightness beyond x $<$ -700 km is not included in this linear fit because this region appears to be more strongly contaminated by instrumental backgrounds than the rest of the re-projected image at certain wavelengths. After removing the background across all values of x we define the plume's Equivalent Width at each y corresponding to an altitude z such that $z = -(250 + y)$ in km as the total integrated brightness in a horizontal slice through the plume \citep{hedman2013observed}. The Equivalent Width is calculated as the sum of the signal across x dimension multiplied with the step width (that is, 100 km). This yields the plume's Equivalent Width for all y between -2000 km and -300 km, or altitudes $z$ ranging from 50 km to 1750 km. This process is carried out on each binned cube for all wavelength channels of each of the 3 dates.

Figure~\ref{fig:wavelength filter} shows example plume spectra from 0.9 $\mu$m to 4.0 $\mu$m obtained from cubes obtained close to peak of Enceladus' activity and at low altitudes. Each of these spectra has a clear dip around 3 $\mu$m corresponding to the fundamental water-ice absorption band \citep{2009ApJ...701.1347M}.
While the signal-to-noise of these spectra is reasonably good, for other orbital phases and altitudes it is still rather low and so to better quantify relevant spectral trends we further average these spectra over wavelength. 

For this analysis we focus primarily on wavelength ranges outside the water-ice absorption band. Hence we first apply a median filter to find outliers before 2.5 $\mu$m and after 3.3 $\mu$m range. More specifically, we compute the median of the signal before 2.5 $\mu$m and flag outliers beyond the range of 1.5 times the standard deviation of all the points in this range, and then do the same for the signals beyond 3.3 $\mu$m range. This threshold value of 1.5$\sigma$ was chosen because it was found to remove clear outliers based on visual inspection of selected spectra. After flagging these general outliers, we compute the average and error on the mean signal in bins of 8 wavelength channels each. These averages and errors are computed after excluding both the outliers flagged previously and any data points that are beyond the 2$\sigma$ range from the median of the 8 wavelength channels that are being averaged together. This leaves us with 23 averaged wavelength values, which are shown as the connected lines in Figure~\ref{fig:wavelength filter}. Finally, we compute the weighted average Equivalent Width over four wavelengths each and obtain the plume's output centered at 1.2 $\mu$m, 1.7 $\mu$m, 2.2 $\mu$m and 3.7 $\mu$m. This last averaging step improves the signal to noise ratio considerably, and is particularly useful for the calculations of overall plume output and typical launch velocity described in the next subsection.

\subsection{Quantifying trends with altitude and orbital phase/time} \label{subsec:vel}
In order to better quantify the trends in the brightness with time and altitude, we use the same basic parametrization as \citet{hedman2013observed}.
That work defined a parameter $Z = [z/(r_E+z)]^{1/2}$ where $r_E$ = 250 km is the radius of Enceladus and $z$ is the plume's altitude. This parameter is useful because for low-optical-depth systems like the plume, it is reasonable to assume that Enceladus’ gravity is the dominant force acting on the particles and the particle and gas density are so low that the particles follow purely ballistic trajectories. In this limit, the particle launch velocity $v$ is directly related to the altitude it reaches $z$:
\begin{equation} \label{eq:1}
v = v_{esc}\Big[\frac{z}{r_E + z}\Big]^{1/2} = v_{esc}*Z
\end{equation}
where $v_{esc}$ = 240 $ms^{-1}$ is the escape velocity on Enceladus. Thus for a population of particles, trends in the plume's brightness with $Z$ reflect trends in the particles' launch velocity. \citet{hedman2013observed} found that at wavelengths around 1 $\mu$m the relationship between Equivalent Width and $Z$ was roughly linear with a negative slope. 

\begin{figure}[t]
\plotone{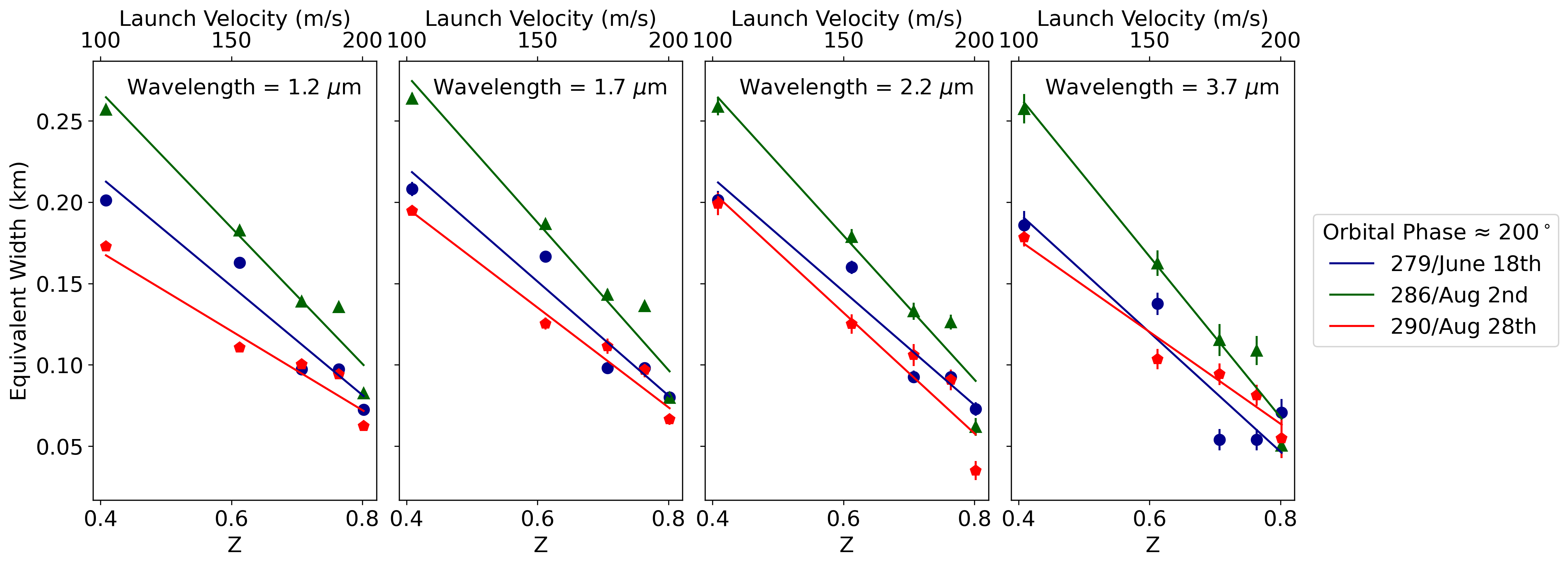}
\caption{The  integrated brightness of the plume (Equivalent Width or EW) as a function of the parameter $Z$ (=$[z/(r_E+z)]^{1/2}$ where $r_E$ = 250 km is the radius of Enceladus and $z$ the plume's altitude) and launch velocity $v$ (calculated using equation~\ref{eq:1}). The Equivalent Width for the three different dates in 2017 - June 18th, Aug 2nd and Aug 28th are plotted (in blue, green and red respectively) at similar orbital phase value of $\simeq$ 200$^{\circ}$ and 4 different wavelengths. The Equivalent Width is fit to a linear function of $Z$. Note the brightness of the plume is higher for the observation on Aug 2nd (in green) than on the other dates.
\label{fig:linear fit}}
\end{figure}

Figure~\ref{fig:linear fit} shows that the approximately linear relationship between the plume's integrated brightness  and the $Z$ parameter observed by \citet{hedman2013observed} also holds for the VIMS observations in 2017 over all the observed wavelengths. We therefore fit a linear trend to the plume's EW profile versus $Z$ in Figure~\ref{fig:linear fit} at altitudes between 50 km and 450 km i.e., $Z$ between 0.41 and 0.8 and launch velocity $v$ between 100 $ms^{-1}$ and 200 $ms^{-1}$. 

The parameters for this linear fit are then used to calculate two quantities. One parameter is the Equivalent Width (a measure of the plume's total brightness) at a reference altitude of z = 85 km ($Z$ = 0.5; $v$ = 120 $ms^{-1}$) calculated through interpolation using the slope and the y-intercept, while the other is a critical velocity $v_c$, which is the value of $v$ where the linear trend in the Equivalent Width would pass through zero.

Note that the critical velocity defined above involves  an extrapolation of the linear fit into regions where it is not necessarily appropriate since this parameter often exceeds  the escape velocity of Enceladus $v_{esc}$. Hence, we instead use this parameter to compute a quantity called the typical launch velocity $v_{typical}$. This corresponds to the weighted average of launch velocities of the particles visible between altitudes of 50 km and 450 km:

\begin{equation} \label{eq:2}
v_{typical} = \displaystyle \frac{\int_{v_{min}}^{v_{max}} n(v) v \,dv }{\int_{v_{min}}^{v_{max}} n(v) \,dv}
\end{equation}
where $v_{min}$ = 100 $ms^{-1}$ and $v_{max}$ = 200 $ms^{-1}$ are the minimum and maximum launch velocity of particles in the range of altitude 50 km to 450 km, and $n(v)$ is the launch velocity distribution of the particles.
For this analysis, we assume $n(v) \propto (1 - v/v_{c})$, consistent with the observed linear trend between Equivalent Width and launch velocity shown in Figure~\ref{fig:linear fit}. Using these values, Equation~\ref{eq:2} can be reduced to:

\begin{equation} \label{eq:3}
v_{typical} = \frac{\Bigg[\displaystyle \frac{v_{max}^2}{2} - \frac{v_{max}^3}{3v_{c}}\Bigg] - \Bigg[\frac{v_{min}^2}{2} - \frac{v_{min}^3}{3v_{c}}\Bigg]}{\Bigg[\displaystyle v_{max} - \frac{v_{max}^2}{2v_{c}}\Bigg] - \Bigg[v_{min} - \frac{v_{min}^2}{2v_{c}}\Bigg]}
\end{equation}

Using the above equations, the typical launch velocity is calculated for each binned cube.
Since the trends among the different panels in Figure~\ref{fig:linear fit} are nearly the same, the typical launch velocity should not depend strongly on wavelength.

\section{Results}
\label{sec:results}

\begin{figure}[t]
\plotone{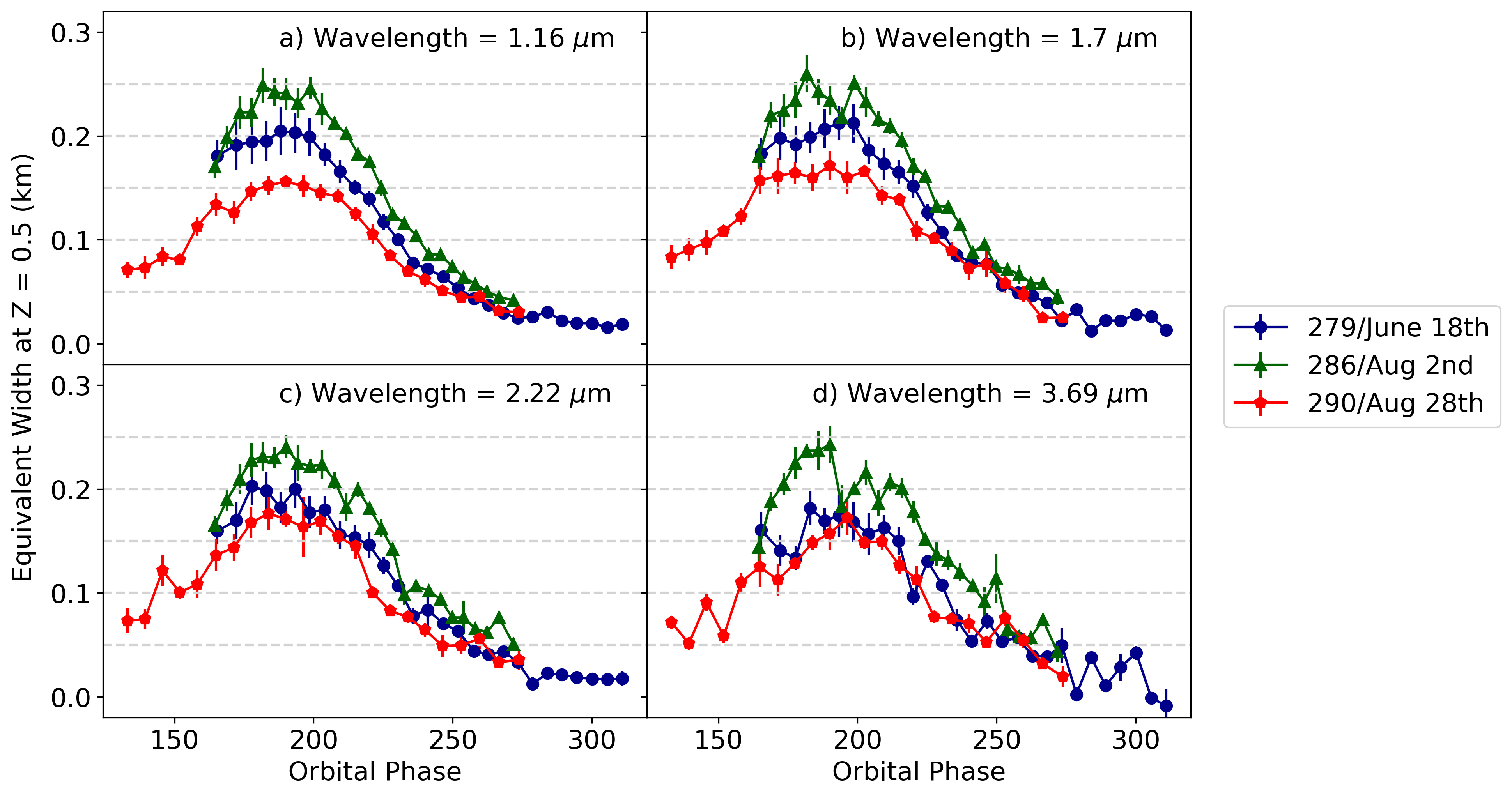}
\caption{The plume's Equivalent Width at $Z$ = 0.5 (altitude $z$ = 85 km) 
as a function of orbital phase for the three dates and four different wavelength values. The plume's maximum brightness around the orbital phase of 180$^\circ$ is consistent at all wavelengths. Additionally, note that the brightness of the plume was higher on Aug 2nd than it was on June 18th or Aug 28th in 2017 at all wavelengths.
\label{fig:EW across orbits}}
\end{figure}
\begin{figure}[t]
\plotone{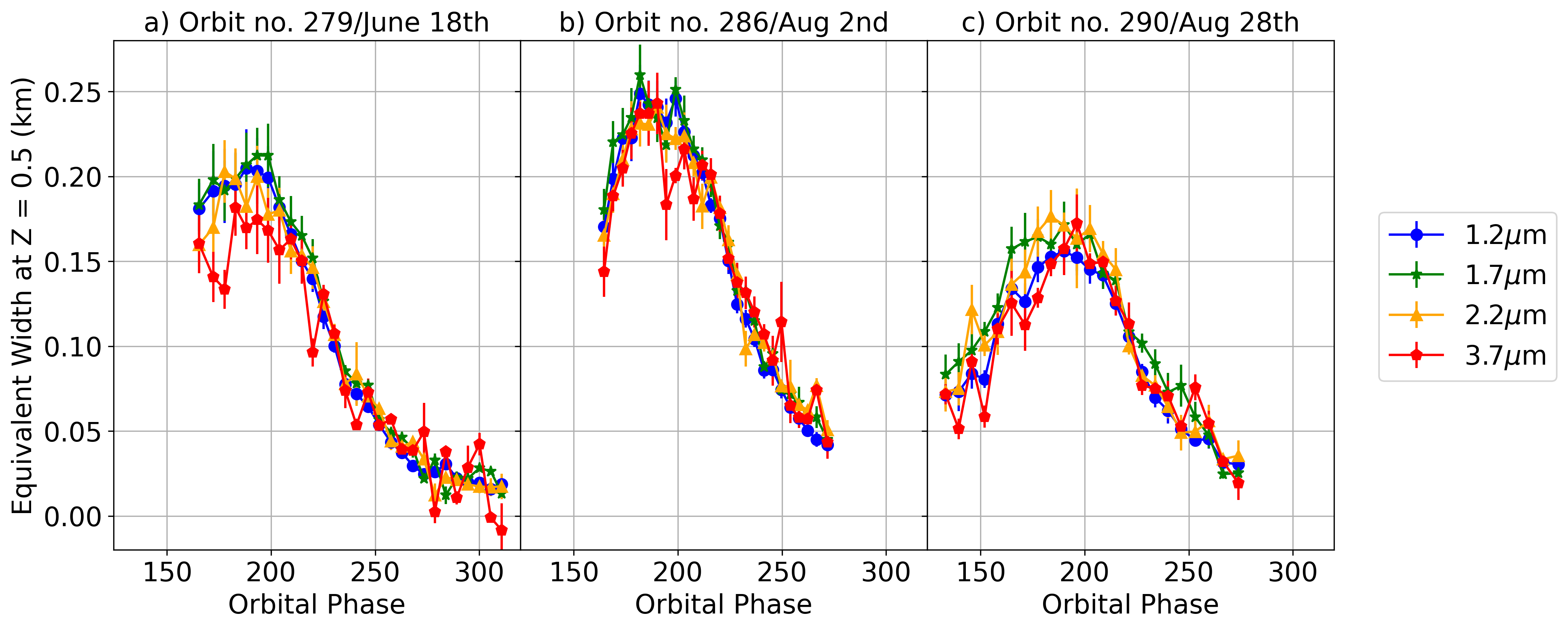}
\caption{The plume's Equivalent Width at $Z$ = 0.5 (altitude $z$ = 85 km, same as Figure~\ref{fig:EW across orbits}) as a function of orbital phase, grouped by the Orbit no./Date instead of wavelength. Note that on Aug 2nd the brightness variations are nearly identical at all wavelengths, spectral trends in the plume's brightness can be seen in both the June 18 and Aug 28 data at orbital phases below 180$^\circ$. 
\label{fig:EW across wavelengths}}
\end{figure}
\begin{figure}[t]
\plotone{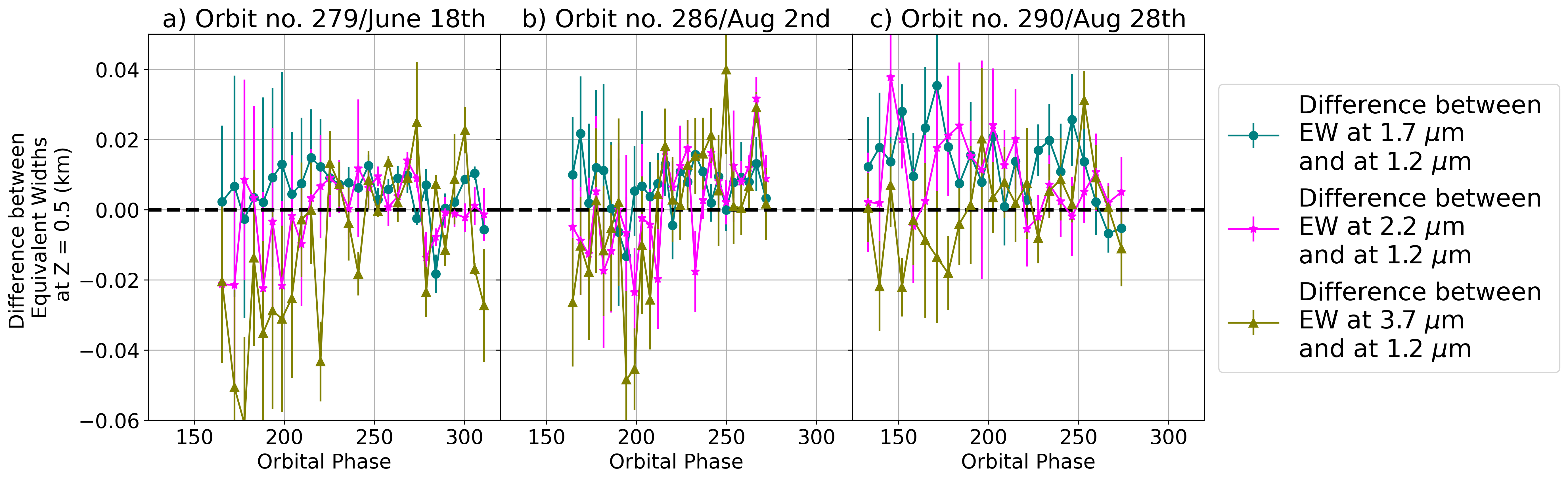}
\caption{The difference in plume's Equivalent Width at different wavelengths and at $Z$ = 0.5 (altitude $z$ = 85 km) as a function of orbital phase, grouped by the Orbit no./Date. Note that on Aug 2nd the brightness variations are nearly identical at all wavelengths, spectral trends in the plume's brightness can be seen in both the June 18 and Aug 28 data at orbital phases below 180$^\circ$.}
\label{fig:Difference in EW}
\end{figure}

Figures~\ref{fig:EW across orbits} and ~\ref{fig:EW across wavelengths} show the plume's Equivalent Width at $Z$ = 0.5 as a function of orbital phase for all three dates (June 18th, Aug 2nd and Aug 28th) at 4 different wavelengths. In addition, Figure~\ref{fig:Difference in EW} shows the differences in the  Equivalent Width at different wavelengths for each of the three dates. The Equivalent Width parameter provides an estimate of the plume's overall brightness and therefore the moon's overall activity level. All three observations covered the range of orbital phase 160$^\circ$ to 270$^\circ$. The plume's maximum brightness is consistently around the orbital phase of 180$^\circ$ for all these observations, regardless of wavelength. This confirms that the plume's ice-particle output is highest when the satellite is furthest from Saturn, consistent with prior analysis of ISS and VIMS plume observations also shown in Figure~\ref{fig:combined} \citep{hedman2013observed, nimmo2014tidally, ingersoll2017decadal}.

Closer comparisons of these data with the ISS data reported in \citet{ingersoll2020time} not only confirm some of the variations documented in that work, but also highlight novel spectral trends.
In Figure~\ref{fig:EW across orbits} at wavelengths of 1.2 $\mu$m the trends with time and orbital phase among the observations are similar to those seen at visible wavelengths \citep{ingersoll2020time}, with the plume being brighter on August 2nd than it was on June 18th and August 28th, indicating the overall activity level in the plume rose and fell during the 10-week interval of these observations. However, the relative brightness of the plume on June 18th and August 28th also varies with wavelength (see also Figure~\ref{fig:EW across wavelengths} and~\ref{fig:Difference in EW}). At short wavelengths of 1.2 $\mu$m and 1.7 $\mu$m shown in panels a) and b) of Figure~\ref{fig:EW across orbits}, the plume is significantly brighter on June 18th (in blue) than it was on August 28th (in red), which is consistent with prior analysis of the imaging data \citep{ingersoll2020time}. However at a longer wavelengths of 3.7 $\mu$m in panel d) of Figure~\ref{fig:EW across orbits} the plume's brightness on these two dates (in blue and red) are nearly identical. See Section~\ref{sec:discuss}  for further details on this variation.

\begin{figure}[t]
\plotone{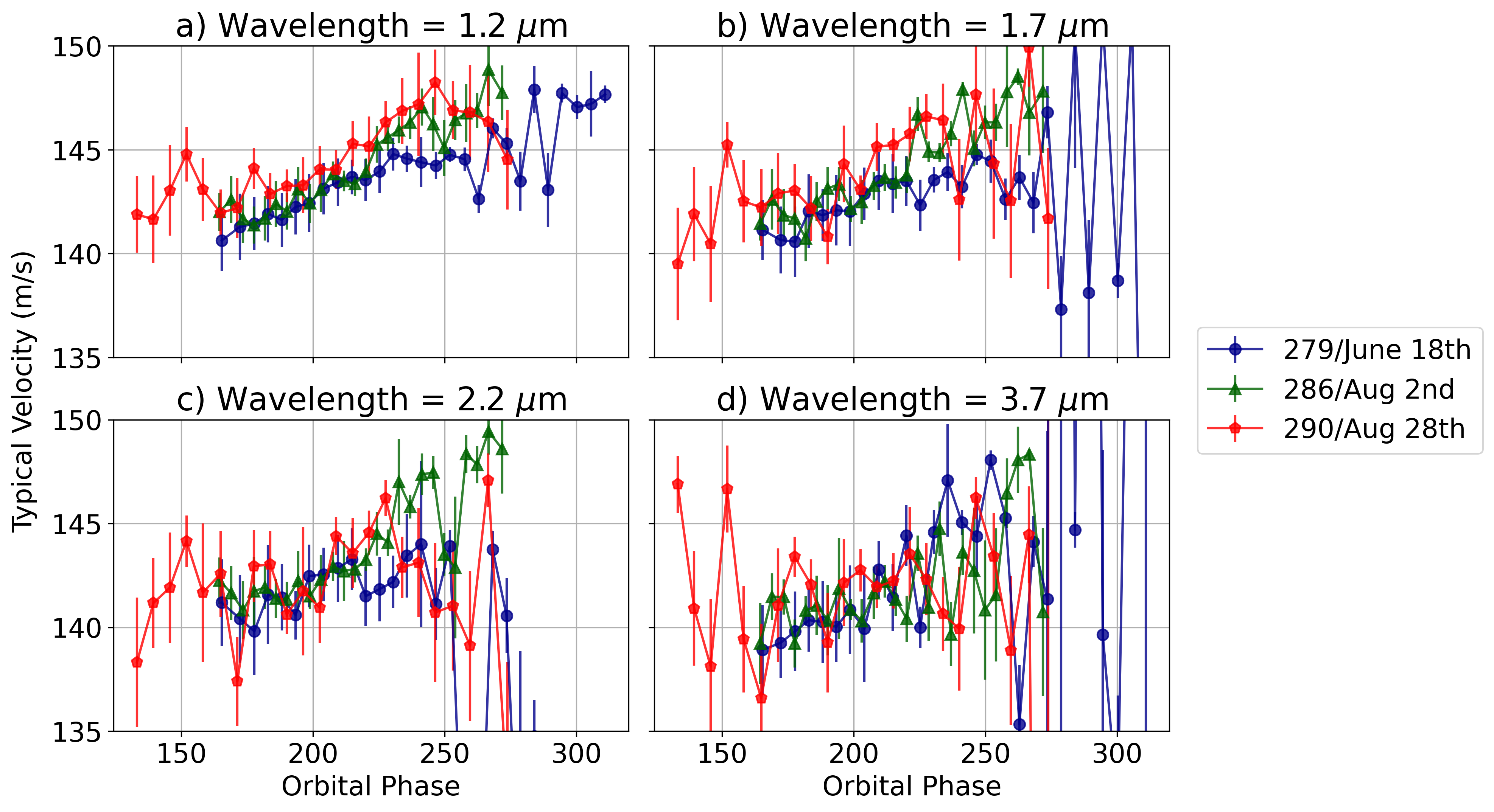}
\caption{Typical launch velocity (in $ms^{-1}$) for three different orbits derived using equation~\ref{eq:3} as a function of orbital phase. Note that the typical launch velocity is higher on Aug 28th than June 18th or Aug 2nd at lower wavelengths of 1.2 $\mu$m. The typical launch velocity also generally increases with orbital phase. Also note the variations in the typical launch velocity with wavelength are relatively small.
\label{fig:Typical velocity}}
\end{figure}

Figure~\ref{fig:Typical velocity} shows the typical launch velocity of the plume particles derived from the same linear fits described above as a function of orbital phase for the same four average wavelength values and three observation dates. The typical launch velocity plotted across orbital phase in Figure~\ref{fig:Typical velocity} are averaged using the inverse of variance as weights. The weighted averages of these typical launch velocities across four wavelength ranges and 5 orbital phase values are also provided in Table~\ref{tab:vel}.
Unlike the dramatic variations seen in the plume's brightness, this parameter depends much less on orbital phase, date or wavelength.
In general, the typical launch velocity increases with increasing orbital phase between 160$^\circ$ and 230$^\circ$. 
The spectral variations in the typical launch velocity are subtle, with the values at 3.7$\mu$m being only about 10 m s$^{-1}$ less than the values at 1.2 $\mu$m (see Table~\ref{tab:vel}).
This is consistent with Figure~\ref{fig:linear fit} where the linear trends for each Orbit no./Date are similar for all the different wavelengths.
Also note that at 1.2 $\mu$m the typical launch velocity is slightly higher on Aug 28th than on June 18th or Aug 2nd. 

\begin{deluxetable*}{ccccc}
\tablenum{2}
\tablecaption{Typical launch velocity for the three orbits being studied in this paper across orbital phase and wavelength.\label{tab:vel}}
\tablewidth{0pt}
\tablehead{
 & \colhead{} & \multicolumn{3}{c}{Typical launch Velocity ($ms^{-1}$)}\\ \cline{3-5}
 \colhead{Wavelength} & \colhead{Orbital Phase} & \colhead{279} & \colhead{286} & \colhead{290}\\
\colhead{($\mu$m)} & \colhead{($^\circ$)} & \colhead{June 18th} & \colhead{Aug 2nd} & \colhead{Aug 28th}
}
\startdata
    \multirow{3}{5em}{0.95 - 1.37}  &  174.6 &                   141.3 $\pm$ 0.7  &  141.7 $\pm$ 0.5  & 142.9 $\pm$ 0.6 \\
     & 196.0  &  142.4 $\pm$ 0.7  &  142.6 $\pm$ 0.5 &                   143.5 $\pm$ 0.6 \\
     &  217.4 &  143.7 $\pm$ 0.5 &  143.9 $\pm$ 0.3 &                   145.1 $\pm$ 0.5 \\
     &  238.4 &  144.5 $\pm$ 0.4 &  146.2 $\pm$ 0.4 &                   147.5 $\pm$ 1.0 \\
     &  260.2  &  144.8 $\pm$ 0.2 &  146.9 $\pm$ 0.5 &                   146.8 $\pm$ 1.1 \\ \hline
    \multirow{3}{5em}{1.50 - 1.89} &  174.6 &                   141.1 $\pm$ 0.8 &   141.4 $\pm$ 0.5 &                   142.6 $\pm$ 0.8 \\
     &  196.0 &  142.2 $\pm$ 0.7 &   142.7 $\pm$ 0.4 &                   142.8 $\pm$ 0.6 \\
     &   217.4 &  143.1 $\pm$ 0.7 &  144.2 $\pm$ 0.4 &                   145.6 $\pm$ 0.5 \\
     &  238.4 &  144.1 $\pm$ 0.3 &   146.5 $\pm$ 0.2 &                   145.9 $\pm$ 1.3 \\
     &  260.2 &   143.4 $\pm$ 0.6 &   148.1 $\pm$ 0.4 &                   147.7 $\pm$ 1.5 \\ \hline
    \multirow{3}{5em}{2.01 - 2.41} & 174.6 &                   140.7 $\pm$ 1.1 &  141.8 $\pm$ 0.6 & 141.9 $\pm$ 0.9 \\
     &   196.0 &  141.7 $\pm$ 0.7 &  141.6 $\pm$ 0.5 &                   140.8 $\pm$ 0.8 \\
     &  217.4 & 142.3 $\pm$ 0.8 &  143.2 $\pm$ 0.3 &                   145.0 $\pm$ 0.5 \\
     &  238.4 &  142.2 $\pm$ 0.9 &  145.8 $\pm$ 0.4 &                   142.7 $\pm$ 1.2 \\
     &  260.2 &  142.6 $\pm$ 0.6 &   148.3 $\pm$ 0.5 &                   145.5 $\pm$ 1.1 \\ \hline
    \multirow{3}{5em}{3.49 - 3.90} &  174.6 &                   139.7 $\pm$ 0.9 &   140.7 $\pm$ 0.5 &                   142.5 $\pm$ 0.7 \\
     &  196.0 &   140.3 $\pm$ 1.0 &   140.8 $\pm$ 0.4 &                   142.2 $\pm$ 0.9 \\
     & 217.4 &   141.7 $\pm$ 0.6 &  141.9 $\pm$ 0.5 &                   142.2 $\pm$ 0.7 \\
     &    238.4 &    145.0 $\pm$ 0.5 &    142.0 $\pm$ 0.8 &   144.5 $\pm$ 0.8 \\
     &   260.2 &   146.1 $\pm$ 0.3 &  148.2 $\pm$ 0.3 &                   143.1 $\pm$ 1.4 \\ \hline
\enddata
\end{deluxetable*}

\section{Discussion}
\label{sec:discuss}
Figures~\ref{fig:EW across orbits} - \ref{fig:Difference in EW} show several interesting trends with orbital phase, time, and particularly wavelength. Since the wavelength trends are the most unique aspect of these new VIMS data, we will focus primarily on these aspects of the data here. In general, particles of different sizes scatter different wavelengths of light with different efficiencies, so the spectral trends provide information about trends with particle size. Detailed modeling of these spectra  will be the subject of a future work, but we can already highlight some interesting trends with wavelength that likely reflect trends with particle size. VIMS data are observed at high phase angles, where particles scatter light most efficiently at wavelength comparable to the particle radius \citep{van1957light, hedman2009spectral}. Hence larger particles contribute more to the plume's brightness at longer wavelengths. In other words, larger particles tend to produce a spectrum with a redder slope.

Variations in the plume spectra indicative of variations in the particle size distribution are most easily seen in Figure~\ref{fig:Difference in EW}. One June 18th, the olive green curve is more negative (brightness decreases from 1.7 $\mu$m to 3.7 $\mu$m) hinting at a higher number of small particles in the plume on that date. While on Aug 28th both the magenta and teal curves are above the zero line (brightness increases from 1.2 to 1.7 $\mu$m and 2.2 $\mu$m). This variation is also visible in the spectra shown in Figure~\ref{fig:wavelength filter}, where the Aug 28th data shows a redder slope than the earlier data on June 18th and Aug 2nd. This indicates that the plume contained a higher fraction of larger particles on Aug 28th than on June 18th. 

These changes in plume's spectra may provide additional clues about what happened to produce the brightness changes across these three dates. One potential explanation for these changes is that they reflect localized sources turning on and off. Individual jets have been observed to turn on and off over time scales that are not explicable by simple tidal models \citep{porco2014geysers, 2017DPS....4920702S, 2020DPS....5221505S, ingersoll2020time}. The maximum in plume brightness might be due to a highly collimated jet only seen on Aug 2nd \citep{ingersoll2020time}. However, this change in plume activity could also be explained by subsequent opening of new channels or choking of conduits by ice deposition in the near-surface \citep{spencer2018plume, ingersoll2010subsurface} 
Changes in the particle size distribution as a whole from one month to another could shed light on the cause of this stochastic variability in the plume such as how these variations reflect changes in vent conditions. We plan to further explore the particle size variations in the plume using Mie scattering in our future work.

Figure~\ref{fig:Typical velocity} shows that for all three orbits, the typical launch velocity of particles increases with orbital phase after the satellite passes the plume maxima near apocenter. This increase in the velocity with orbital phase holds true for all wavelengths and is consistent with prior results \citep{hedman2013observed,ingersoll2020time}. 
This suggests an inverse relation between the particle mass flux and ejection velocity at least in the region beyond the apoapsis. 
Another key observation is the typical launch velocity is higher on Aug 28th than on June 18th and Aug 2nd at wavelengths of 1.2 $\mu$m while the particle mass flux reflected by the Equivalent Width in Figure~\ref{fig:EW across orbits} is lowest. This further alludes to a complementary change in particle mass flux and typical launch velocity at least at lower wavelength of 1.2 $\mu$m. One possible explanation for this is the narrowing of vents due to tidal stresses as the satellite cross its apocenter might increase the flow speed while decreasing the mass flux \citep[but see ][for potential complications with this idea]{nimmo2014tidally}.

Interestingly, the launch velocity of particles in Table~\ref{tab:vel} and Figure~\ref{fig:Typical velocity} shows only a slight decrease as wavelength increases. This is surprising because previously published models by \citet{schmidt2008slow} predict substantial variations of launch velocity with particle size. According to these models, repeated collisions with the walls of the conduit reduce the particle velocity relative to the gas. 
\citet{degruyter2011cryoclastic} also modeled the particle acceleration within the conduit and their ballistic transport once they exited the vent using the gas flow model of \citet{ingersoll2010subsurface} and the \citet{schmidt2008slow}'s collision model and similarly found that larger particles achieve lower exit speeds. Using \citet{schmidt2008slow}'s collision model for a gas density of 4.85 $g m^{-3}$ and gas speed of 500 $m s^{-1}$ and a collision length of 0.1 m, a large fractional reduction of 0.98 is expected in the typical particle velocity as size increases from 1.2 $\mu m$ to 3.7 $\mu m$. By contrast, the typical launch velocities in Figure~\ref{fig:Typical velocity} shows a fractional reduction of only 0.01 - 0.16 as wavelength increases from 1.2 $\mu m$ to 3.7 $\mu m$ which is 6 times lower than predicted by previous models of particle velocities. While the plume's brightness at wavelength is due to particles with a range of sizes, this is still a notable difference.

This finding is also consistent with some of the earlier spectral analysis of the Enceladus plume. \citet{hedman2009spectral} determined the relative number of particles of radii 1, 2 and 3 $\mu m$ versus height in early VIMS plume observations, which were in turn converted into velocity distribution of particles. Data obtained at orbital phases around 90$^\circ$ - 120$^\circ$ indicated that the number density of larger particles of radius 3 $\mu m$ falls faster than the smaller particles of radius 1 $\mu m$ implying a typical lower launch velocity for 3 $\mu m$ particles which was roughly consistent with the \citet{schmidt2008slow}. However, the velocity distribution for the data obtained closest to apopasis in Figure 6 of that paper is not significantly steeper for larger particle sizes for most velocities, indicating the typical launch velocity of larger particles does not decrease steeply, which is more consistent with these observations.

One possible explanation for this surprisingly subtle reduction in particle velocity with increasing particle size is that the vent parameters are different from what was assumed in \citet{schmidt2008slow} model. Changing these parameters such that the critical grain radius is larger could result in a less steep dependence of ejection velocity on grain sizes in the range from sub-microns to a few microns \citep{2014AGUFM.P51F..05S, postberg2009sodium, postberg2011salt}.
Another possible explanation is that particle-particle interactions at the vent are more common than previously thought. Unlike collisions with the walls of the conduit (that produce trends in particle velocity with size \citep{schmidt2008slow}), particle-particle collisions  would cause grains of different sizes to have similar velocity distributions. 
Both particle-wall and particle-particle collisions are most important near the vent where the gas density rapidly declines, and the particles become partially decoupled from the gas (so that they are launched at much lower velocities than the gas) \citep{goldstein2018enceladus}.
Assuming a gas density and a particle density of $10^{23}$ m$^{-3}$ and $2.1 \times 10^{10}$  m$^{-3}$ respectively \citep{yeoh2017constraining} at the end of the conduit, the mean free path for particle-gas collision is of the order of $10^{-13}$ m while for particle-particle collision is 1 $m$. The mean free path for particle-particle collisions is comparable to the measured vent sizes on the south pole of Enceladus. \citet{goguen2013temperature} estimated a fissure width of 9 m based on near-IR thermal emission spectra acquired by VIMS and \citet{yeoh2015understanding} estimated vent diameters of up to 2.8 m. Hence particle-particle interactions could occur at a high enough rate to affect particle velocities, which has not been taken into account in the current models. 

We plan to model the observed spectral trends to obtain quantitative constraints on the particle size distribution at different altitudes and times. 
This information should further information about conditions within the vents.


\section*{Acknowledgements}
We would like to acknowledge J.N. Cuzzi for helpful conversations. This particular work was supported by a Cassini Data Analysis Program Grant 80NSSC18K1071.

\bibliography{sample631}{}
\bibliographystyle{aasjournal}

\appendix
\section{Cubes excluded from analysis}
\label{cubes}

The following cubes were removed from the data on June 18th/Orbit no. 279: CM$\_$1876443559$\_$1, CM$\_$1876443614$\_$1, CM$\_$1876443669$\_$1, CM$\_$1876443724$\_$1, CM$\_$1876445164$\_$1, CM$\_$1876446029$\_$1, CM$\_$1876446202$\_$1, CM$\_$1876446375$\_$1, CM$\_$1876457449$\_$1, CM$\_$1876470103$\_$1, CM$\_$1876474082$\_$1, CM$\_$1876488640$\_$1, CM$\_$1876493658$\_$1, CM$\_$1876493831$\_$1, CM$\_$1876494004$\_$1, CM$\_$1876494177$\_$1, CM$\_$1876494350$\_$1, CM$\_$1876494523$\_$1, CM$\_$1876494696$\_$1, CM$\_$1876494869$\_$1, CM$\_$1876495042$\_$1, CM$\_$1876495215$\_$1. 

The following cubes are removed during background removal from the data on Aug 2nd/Orbit no. 286: CM$\_$1880355922$\_$1, CM$\_$1880358137$\_$1, CM$\_$1880366306$\_$1, CM$\_$1880369490$\_$1, CM$\_$1880380704$\_$1, CM$\_$1880390533$\_$6, CM$\_$1880391952$\_$6, CM$\_$1880392472$\_$6, CM$\_$1880392610$\_$6, CM$\_$1880392818$\_$6, CM$\_$1880392852$\_$6, CM$\_$1880393025$\_$6, CM$\_$1880393233$\_$6, CM$\_$1880393337$\_$6, CM$\_$1880393441$\_$6, CM$\_$1880393579$\_$6, CM$\_$1880393718$\_$1, CM$\_$1880394687$\_$6, CM$\_$1880394791$\_$6, CM$\_$1880395102$\_$6, CM$\_$1880395241$\_$6, CM$\_$1880395794$\_$6, CM$\_$1880395967$\_$6, CM$\_$1880396071$\_$6. 

The following cubes are removed during background removal from the data on Aug 28th/Orbit no. 290: CM$\_$1882608048$\_$5, CM$\_$1882637326$\_$1, CM$\_$1882640856$\_$1, CM$\_$1882644178$\_$1, CM$\_$1882644386$\_$1, CM$\_$1882645009$\_$1, CM$\_$1882646047$\_$1, CM$\_$1882646255$\_$1, CM$\_$1882646463$\_$1, CM$\_$1882646670$\_$1, CM$\_$1882646878$\_$5.



\end{document}